\documentclass[12pt]{iopart}
\usepackage{graphicx}

    \pdfoutput=1

\newcommand{\Py}{Ni$_{81}$Fe$_{19}$}

\begin{document}

\title{Modification of the thermal spin-wave spectrum in a \Py\ stripe by a domain wall}

\author{C.W.~Sandweg, S.J.~Hermsdoerfer, H.~Schultheiss, S. Sch\"afer, B.~Leven, and B.~Hillebrands}

\address{Fachbereich Physik and Landesforschungszentrum OPTIMAS,
Technische Universit\"at Kaiserslautern, Erwin-Schr\"odinger-Stra{\ss}e 56, 67663 Kaiserslautern, Germany}
\ead{leven@physik.uni-kl.de}
\begin{abstract}
The thermal spin-wave distribution in a \Py\ stripe with an
asymmetric transverse domain wall has been investigated using
Brillouin light scattering microscopy. Clear evidence has been
found that the existence of the domain wall influences the
spin-wave distribution of the thermal modes. The thermal spin-wave
modes are quantized due to the confinement in radial direction.
They vanish near the domain wall and a new mode evolves inside
this complex domain wall structure. This effect is attributed to a
change of the effective internal field in the domain wall region.
The experimental results agree well with static and dynamic
micromagnetic simulations.

\end{abstract}

\maketitle

\section{Introduction}
The research field of spin waves in confined magnetic structures
with controlled inhomogeneous internal field is of fundamental
interest for understanding magnetization dynamics (see e.g.
\cite{Demokritov06,Tamaru,Crawford}). Such an inhomogeneous
internal field can be caused by the finite size of the magnetic
object due to stray field effects at the boundaries
\cite{Demokritov06,Kostylev07,Bayer,Roussigne} and/or by a
magnetic domain structure
\cite{Crawford,Bailleul2,Bailleul3,Schultheiss08}. Several
investigations of spin-wave spectra in finite elements without and
with domain structure have been reported (see e.g. References
\cite{Schultheiss08} to \cite{Bailleul1}). Here, we report on the
modification of the thermal spin-wave distribution by a domain
wall. As one main result we report the finding of a spin-wave mode
localized to the domain wall region caused by the spin-wave
potential well generated by the domain wall. To approach the
problem, two main obstacles have to be overcome: firstly and most
importantly, a non-destructive measurement technique capable to
resolve the dynamic properties with high spatial resolution for
resolving the domain pattern is needed. Secondly, domain walls
must be created in a reliable and reproducible way.

Brillouin light scattering spectroscopy using a scanning microfocus
sample stage, hereafter referred to as BLS microscopy, is by now a
well established technique which meets the experimental requirements
regarding imaging of dynamic magnetic properties
\cite{Schultheiss08,Ando07,Demidov05,Demokritov04}. A spatial
resolution of 250\,nm has been achieved with sensitivity down to
thermal spin-wave excitations. This microscopy technique opens the
door to investigate spin waves in a narrow \Py\ stripe with a
well-defined domain wall.

\section{Sample design and experimental setup}
\begin{figure}\center
\includegraphics[width=0.6\columnwidth]{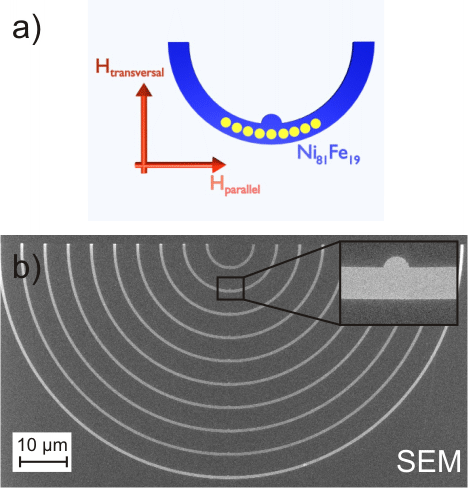}
\caption{\label{setup}(color online) a) Schematic view of the
sample design and data acquisition procedure. Measurements have
been carried out along the positions marked by the bright spots.
The externally applied magnetic fields are shown as
H$_{\mathrm{transversal}}$ and H$_{\mathrm{parallel}}$. b)
Scanning electron micrograph of the \Py\ semi-circles each
comprising an anti-notch as domain wall pinning site.}
\end{figure}

The structure design of curved wires follows an idea presented by
Saitoh \cite{Saitoh} and has also been discussed in \cite{Chappert}
as the so-called domain wall pendulum. A semi-circular \Py\
structure contains a circular anti-notch located at the pole of the
semi-circle acting as a pinning site for the domain wall (see
Fig.~\ref{setup}). In particular, the chosen semi-circular sample
design allows for a defined nucleation and annihilation of a single
domain wall in the vicinity of the anti-notch by applying a magnetic
field in transversal or parallel direction, respectively, and
subsequent relaxation of the magnetization to remanence.

The samples have been produced using a combination of molecular
beam epitaxy and electron beam lithography employing a standard
lift-off process. The \Py\ films have a thickness of 10\,nm. The
\Py\ structures have been prepared using a 120\,nm thick PMMA
resist layer (polymethylmethacrylate, molecular weight 950\,K,
solid fraction of 4\%) spun onto a thermally oxidized Silicon
substrate (500\,$\mu$m Silicon covered with 100\,nm Silicon
oxide). The radii of the structures vary between 5\,$\mu$m and
50\,$\mu$m in steps of 5\,$\mu$m. The wire width is 500\,nm and
the radius of the anti-notch is 250\,nm revealing a total width of
750\,nm at the anti-notch position. The patterned structures show
an induced transverse anisotropy. A schematic view of one of the
semi-circles is shown in Fig.~\ref{setup}\,a). Here, the \Py\
structure is displayed in dark color. The bright spots indicate
the data acquisition positions used. The magnetic field directions
used to nucleate a domain wall or to saturate the sample are also
defined in Fig.~\ref{setup}\,a). Figure~\ref{setup}\,b) shows a
corresponding scanning electron micrograph of the sample design.

For analyzing the spin-wave modes BLS microscopy was employed. As
described above, this method is a powerful non-destructive tool to
obtain space-resolved information in the GHz frequency regime with
a high lateral resolution. The BLS investigations were carried out
exemplarily for the semi-circle with a radius of 5\,$\mu$m. For
magnetostatic domain analysis and dynamic spin-wave distribution
analysis OOMMF \cite{OOMMF} simulations are presented to support
the interpretation of the experimental results.

\section{Experimental results}
An essential prerequisite for the investigation of the
modifications of spin waves in the \Py\ structures due to the
presence of a domain wall is a well characterized domain wall,
that is reproducibly induced and located near the anti-notch.
Since BLS microscopy does not allow for direct in-situ domain
imaging, the domain analysis experiments have been made in
collaboration with the group of Prof John Chapman, Glasgow
University, using Lorentz microscopy. The results are reported in
\cite{Sandweg}. In summary: employing standard electron
transparent Si$_3$N$_4$/Si substrates the domain structure of the
\Py\ wires was investigated. It was found that the domain wall
induced in the semi-circular structures by applying a defined
field sequence exhibit the character of an asymmetric transverse
domain wall and can reproducibly be located close to the position
of the anti-notch.

\begin{figure}\center
\includegraphics[width=0.6\columnwidth]{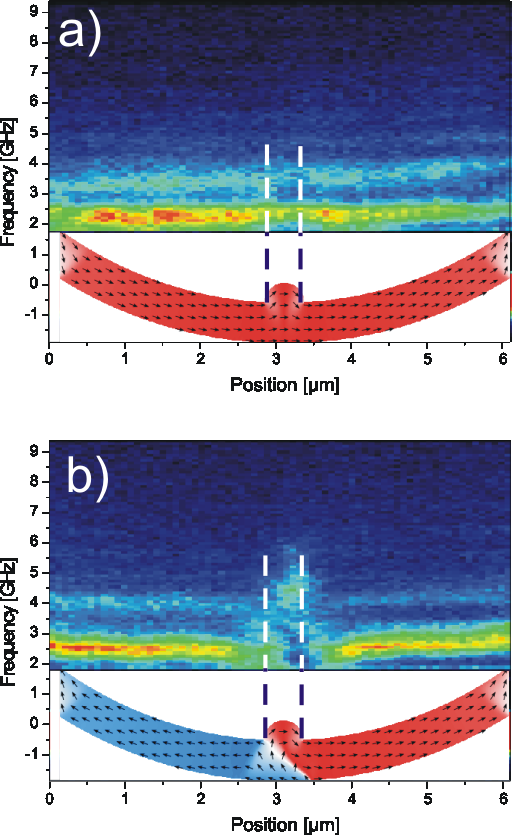}
\caption{\label{BLSmaps}(color online) Intensity map summarizing
the spectra of thermally activated spin waves in the semi-circular
\Py\ structure of radius 5\,$\mu$m without an applied external
field a) in the absence of domain walls, b) in the presence of a
domain wall. The BLS intensity is shown color coded ranging from
dark blue (low intensity) to red (high intensity). Each vertical
line in the two-dimensional map represents a BLS spectrum taken at
the position as indicated on the x-axis. The white dashed lines
show the position of the anti-notch. Insets: Corresponding
magnetization distributions obtained by OOMMF-simulations.}
\end{figure}

To analyze the spin-wave modes present in the \Py\ curved
nanowires the spectra of thermally activated spin waves have been
investigated. The measured BLS spectra are summarized in a color
coded intensity map, where dark blue represents the lowest
spin-wave intensity and red the highest intensity. Each vertical
line in the color map represents a BLS spectrum taken at the
position indicated at the x-axis. The spectra have been taken with
a step size of 0.1\,$\mu$m equidistantly along the central
perimeter of 6.1\,$\mu$m in length in the vicinity of the
anti-notch (see Fig.~\ref{setup}\,a)). First, the sample was
initialized in {\it parallel} direction by saturating the sample
in a field H$_{\mathrm{parallel}}$= 880\,Oe, thus ensuring that no
residual domain walls were present. Subsequently the sample was
relaxed to remanence. Figure~\ref{BLSmaps}\,a) shows the resulting
BLS color map and the corresponding magnetization distribution
obtained by OOMMF simulations \cite{OOMMF}. For the numerical
simulations, the semi-ring with the corresponding radius (inner
radius 5\,$\mu$m, outer radius 5.5\,$\mu$m) has been simulated. To
optimize computation time, only a section of 5.8\,$\mu$m by
1.4\,$\mu$m with a thickness of 10\,nm was taken into account. The
mesh size was 7.5\,nm $\times$ 7.5\,nm $\times$ 10\,nm and the
standard values for \Py\ (exchange stiffness constant
A=1.6$\cdot$10$^{-6}$\,erg/cm, gyromagnetic ratio
$\gamma$=1,76$\times$10$^{-2}$ GHz/Oe, and a damping constant
$\alpha$=0.01) have been used. A saturation magnetization of only
650\,G instead of 860\,G has been chosen to take account of a
heating effect of the sample by the laser spot
\cite{Schultheiss08}.

In this remanence case {\it without} a domain wall
two modes of standing spin waves with frequencies of about 2.4 and
3.4\,GHz can be clearly identified. These modes exhibit mainly the
characteristics of magnetostatic surface waves, the so-called
Damon-Eshbach modes, whose direction of propagation is in-plane
perpendicular to the magnetization of the structure. They are
quantized in transversal direction due to the lateral confinement
of the structure and travel forth and back between the stripe
boundaries. In this regard no influence of the changing boundary
conditions due to the anti-notch is evident in the BLS-spectra.

Next, the change of the thermal spin-wave spectrum due to the
presence of the asymmetric transverse domain wall has been
analyzed. To do so, the sample has been initialized by applying an
external magnetic field in {\it transversal} direction (see
Fig.~\ref{setup}\,a)). After removal of the field an asymmetric
transverse domain wall is nucleated in the structure and pinned in
the vicinity of the anti-notch \cite{Sandweg}. The corresponding
BLS intensity map as well as the according magnetization
distributions obtained by OOMMF-simulations \cite{OOMMF} are shown
in Fig.~\ref{BLSmaps}\,b).

By comparing the intensity maps obtained with and without a domain
wall, obvious differences in the spectra can be determined:
instead of the original modes observed outside the domain wall
with frequencies of 2.4 and 3.4\,GHz a new mode with a frequency
of about 4.8\,GHz arises at the position of the anti-notch whereas
the other modes vanish in this area (see Fig.~\ref{BLSmaps}\,b)).

This behavior can be understood in terms of a change in magnitude
and direction of the effective local internal magnetic field
$\vec{h}_{\mathrm{eff}}$ \cite{Bayer}. Such a local change in the
magnetic field caused by the asymmetric transverse domain wall can
act as a spin-wave potential well supporting a localized mode
\cite{Schultheiss08}. Due to the two-dimensional character of this
problem an analytical approach to calculate the mode frequency is
impractical - instead dynamical micromagnetic simulations have
been carried out for direct comparison.

The dynamic simulations have been performed by applying a weak
out-of-plane Gaussian-shaped magnetic field pulse (amplitude
1\,Oe, pulse width 20\,ps) to the semi-circle at remanence and,
thus, excite the eigenmode spectrum of the magnetic element. These
data have been Fourier-transformed point by point afterwards to
the frequency domain to obtain the spin-wave mode distribution. In
comparison to the experiments the simulations show a slight
frequency shift as the first mode can be observed at 3.3\,GHz
whereas in the experiment the frequency of the first mode is
2.4\,GHz. This can be observed as well for the second mode which
appears in the experiment at 4\,GHz and in the simulations at
4.7\,GHz. This difference can be understood by a difference
between the sample properties and the parameters of the
simulation, e.g. possible variations of the effective stripe width
of the sample caused by fabrication defects which are not taken
into account in the simulation.

\begin{figure}\center
\includegraphics[width=0.8\columnwidth]{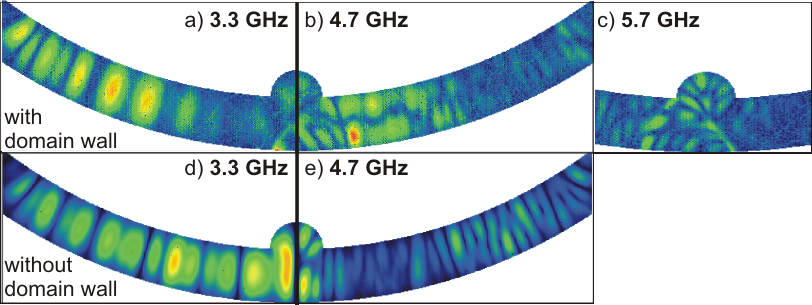}
\caption{\label{FT} (color online) Spatially resolved
Fourier-transformed spin-wave mode distributions obtained by
micromagnetic simulations for a frequency of a)
3.3\,GHz, b) 4.7\,GHz, and c) 5.7\,GHz at remanence.\\
Panel a) shows the first mode along the semi-circle perimeter in
the presence of a domain wall. In Panel b) the node due to the
quantization in radial direction can be clearly seen. Both mode
profiles are strongly disturbed in the region of the anti-notch,
where the domain wall is localized. Panel c) shows the weakly
excited mode in the area of
the anti-notch for high frequencies.\\
Panels d) and e) show the results of the simulation in the absence
of a domain wall but at the same frequencies as before. It can be
seen that the mode structure does not change significantly in the
vicinity of the anti-notch.}
\end{figure}

In general, the experiment and the simulation show a very good
qualitative agreement. Fig.~\ref{FT} shows the first modes in the
structure without an applied external magnetic field. In
Fig.~\ref{FT}a) the first mode without a node can be observed
along the whole semi-circle perimeter, suppressed only in the area
of the domain wall. The second mode shown in Fig.~\ref{FT}b) shows
the typical node in the middle of the stripe due to quantization
of the spin waves in radial direction. For both modes a clear
disturbance is observed in the region of the domain wall.
Fig.~\ref{FT}c) shows a weakly excited mode at higher frequencies,
which is mainly localized in the area of the anti-notch, where the
domain wall is pinned. This mode corresponds to the experimentally
observed mode localized in the domain wall (see
Fig.~\ref{BLSmaps}b)). For comparison, the simulations have been
carried out for the same structure and at remanence but without a
domain wall, i.e. in an uniformly magnetized semi-circle (see
Fig.~\ref{BLSmaps}a)). The results are shown in Fig.~\ref{FT}d)
and e) for the corresponding frequencies. In this configuration no
major change in the mode structure in the vicinity of the
anti-notch can be observed. This result proofs that the pinned
domain wall is the reason for the changes in the mode structure.

\begin{figure}\center
\includegraphics[width=0.8\columnwidth]{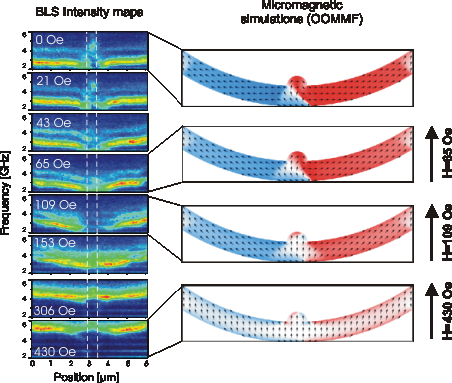}
\caption{\label{BLS_transvH} (color online) Experimental intensity
map summarizing the spectra of thermally activated spin waves in
the semi-circular \Py\ structure of radius 5\,$\mu$m while
applying a transversal field ranging from 0\,Oe to 430\,Oe. The
BLS intensity is shown color coded ranging from dark blue (low
intensity) to red (high intensity). Each vertical line in the
two-dimensional map represents a BLS spectrum taken at the
position as indicated on the x-axis. The white dashed lines show
the position of the anti-notch. Right side: Corresponding
magnetization distributions
obtained by OOMMF-simulations.\\
The intensity maps show the disappearance of the original modes at
the position of the anti-notch and the appearing new mode in this
area. With increasing fields, the modes shift to higher
frequencies and the domain wall is broadening. In the last maps
which correspond to high field values, the domain wall vanished
and only the effect of the increased stripe width at the position
of the anti-notch on the mode frequency can be observed. }
\end{figure}

To further analyze this new mode localized in the vicinity of the
domain wall the behavior of the spin-wave modes under the
influence of an increasing {\it transversal} as well as an
increasing {\it parallel} magnetic field was investigated. First,
the domain wall growth and annihilation expected for an increasing
{\it transversal} magnetic field is discussed. During this
procedure the area the domain wall width is increasing. For a
better comparison between the intensity maps at different fields,
only the BLS frequencies between 2 to 6\,GHz are shown in
Fig~\ref{BLS_transvH}. Additionally, the results of the
corresponding micromagnetic simulations are added for selected
fields. As can be seen from the BLS intensity maps and comparison
with the micromagnetic simulations, the new mode inside the domain
wall is pronounced as long as the asymmetric transverse domain
wall exists. The mode starts to vanish when the domain wall width
is increasing due to the externally applied transversal field. The
asymmetric transverse domain wall loses its character at 109\,Oe.
For further increase of the transversal field the magnetization
follows the external field even in the area of the anti-notch.

\begin{figure}\center
\includegraphics[width=0.8\columnwidth]{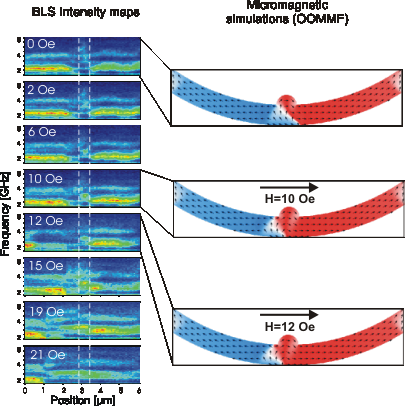}
\caption{\label{BLS_parrH} (color online) BLS intensity maps
summarizing the spectra of thermally activated spin waves in the
semi-circular \Py\ structure of radius 5\,$\mu$m for different
applied parallel fields. For this measurement the domain wall has
been nucleated by applying a transversal field and relaxation to
remanence, afterwards the sample has been rotated by 90\,$^\circ$
and a parallel field has been applied. The BLS intensity is shown
color coded ranging from dark blue (low intensity) to red (high
intensity). Each vertical line in the two-dimensional map
represents a BLS spectrum taken at the position as indicated on
the x-axis. The white dashed lines show the position of the
anti-notch. Right side: Corresponding magnetization distributions
obtained by OOMMF-simulations.\\
The field driven displacement of the domain wall can be seen from
the intensity maps as a change in the mode spectrum. As long as
the domain wall is pinned in the vicinity of the anti-notch the
characteristics of the mode spectrum on the left hand side of the
wall does not change while the field changes. As soon as the wall
is depinned due to the increased external field, also the mode
spectrum on the left hand side changes accordingly}.
\end{figure}

Second, the spin-wave distribution with increasing {\it parallel}
magnetic field was investigated. In this measurement geometry the
asymmetric transverse domain wall is expected to extend and depin
from the pinning site, as has been demonstrated in \cite{Sandweg}.
This time the domain wall was initialized as before but the sample
was rotated afterwards by 90\,$^\circ$ and a parallel, slightly
increasing field has been applied. The results as well as
corresponding micromagnetic simulations are presented in
Fig.~\ref{BLS_parrH}.

As can be seen from the simulations, the domain wall is depinned
already at a field of 10\,Oe and driven to the left by the
external field. This change in the magnetization distribution can
also be observed in the BLS intensity maps. As long as the domain
wall is pinned in the vicinity of the anti-notch the eigenmode
spectrum with the first mode at 2.5\,GHz on the left hand side
does not show a significant change. As soon as the domain wall is
depinned and starts to move in this direction, the modes existing
only outside the domain wall vanish at the position of the wall
which can clearly be seen starting from 12\,Oe. Consequently, the
domain wall displacement also causes a change of the spin-wave
mode profile in the area of the anti-notch, as the domain wall is
not longer localized at this position. The mode localized in the
domain wall disappears at this position as the domain wall moves
and the spin-wave mode profile existing in the semi-circle
perimeter evolves. Thus, the spin-wave mode localized in the
domain wall enables the observation of domain wall displacement by
BLS microscopy. These results are in excellent agreement with
\cite{Sandweg} where the movement of such a domain wall in the
same structure has been observed by means of Lorentz microscopy.

\section{Conclusion}
This article reports on the analysis of the thermal spin-wave mode
distribution in \Py\ semi-circles with defined asymmetric
transverse domain walls employing Brillouin light scattering
microscopy. The spectra of thermally excited spin waves reveal a
clear influence of the presence of the domain wall, which is
nucleated and pinned in the vicinity of the anti-notch of the
structure. Comparing the spin-wave spectra in the absence and
presence of the domain wall it can be observed that the original
spin-wave modes quantized in radial direction vanish in the
vicinity of the domain wall and new modes are formed located only
inside this complex domain wall structure. The experimental
results are confirmed by static and dynamic OOMMF simulations.
Investigating the field dependence of these new modes it could be
proven that the domain wall growth and destruction in a
transversal applied field as well as the depinning behavior in a
parallel applied field can be monitored by BLS microscopy
employing the spatially resolved detection of the mode localized
in the domain wall structure.

\section{Acknowledgment}
The authors thank the Nano+Bio Center (Sandra Wolff, Bert L\"agel, and Christian Dautermann) of the University of
Kaiserslautern for technical support during the sample processing and P. Andreas Beck for thin film deposition.
Financial Support by the DFG within the Priority Programme 1133 "Ultrafast magnetization processes" is gratefully
acknowledged.

\end{document}